\documentclass[12pt]{article}
\usepackage{amsmath,amsfonts, amssymb,graphicx,geometry,authblk,color,soul,comment,hyperref,algorithm,algpseudocode} 
\usepackage{booktabs, array, tabularx, threeparttable}

\DeclareMathOperator*{\argmin}{arg\,min}

\title{Specimen design for material parameter identification using topology optimization}
\author{Adeline Wihardja, Kaushik Bhattacharya }
\date{}

\begin{document}
\maketitle
\begin{abstract}
Constitutive relations close the equations of continuum mechanics, and serve as a surrogate for a material in the design and engineering process. They are often specified in a parameterized form with parameters identified by experiment.   In this paper, we propose a framework for identifying experimental configurations that are maximally informative for material parameter identification. The framework strongly couples modeling and experimentation: the model leverages high-dimensional data from full-field measurements, while the current uncertainty in the model guides the design of future experiments. We formulate this goal by integrating Bayesian optimal experimental design with topology optimization. The Bayesian design criterion quantifies expected information gain, which drives the topology optimization of the specimen geometry.
\end{abstract}


\section{Introduction}
Constitutive relations, which map kinematic quantities (e.g., strain, deformation history) to dynamic quantities (e.g., stress), are essential to close the equations of continuum mechanics \cite{gurtin2010mechanics,holzapfel2000nonlinear} and describe the underlying physics of a material. In practice, identifying constitutive relations of a material involves selecting an appropriate model form, usually informed by the knowledge of the underlying material physics, and estimating its model parameters from experiments. Once calibrated, the model is used to predict material response in more complex engineering scenarios. The quality of these predictions depends on the fidelity of the chosen constitutive model, and thus depends on the quality of its identified parameters.

Unfortunately, identifying constitutive laws and material parameters is challenging since the quantities being mapped by these relations (stresses, strains, internal states) are not directly measurable in an experiment \cite{bonnet2005inverse}. Instead, the most common directly measurable quantities in material testing experiments are boundary loads and boundary displacements, and one has to infer stresses and strains from the measured data to then determine model parameters. Traditional mechanical testing addresses this difficulty by using simple specimen geometries and loading that produce homogeneous deformation in the sample. The measured boundary force provides the corresponding stress state, and repeating this procedure over different loading conditions allows calibration of the constitutive model. This approach has been used extensively for isotropic rubber-like materials \cite{rivlin1951large,treloar1944stress,kawabata1981experimental}. 

Advances in full-field imaging provide an opportunity to obtain substantially richer information from a single experiment. Techniques such as digital image correlation (DIC) provide spatially resolved deformation measurements, allowing heterogeneous deformation fields to be exploited for parameter identification \cite{sutton2009image,yang_augmented_2020,hild,leclerc2009integrated}. These measurements are then combined with methods such as the virtual fields method (VFM) \cite{pierron2012virtual}, finite element model updating (FEMU) \cite{chen2025finite}, or more recently direct image-based constitutive identification methods \cite{wihardja2025constitutive}. In all these established methods, the problem of model discovery in mechanics is often simplified to identifying material parameters within the chosen model form. However, parameter identification remains a challenging inverse problem where material parameters must be inferred from noisy experimental measurements.




As an inverse problem, parameter identification is generally ill-posed and is often posed in a deterministic setting where regularization is introduced to stabilize the inversion. The choice of regularization can be somewhat arbitrary, and selecting an appropriate regularization can require extensive domain knowledge. A Bayesian formulation provides an alternative probabilistic interpretation, where modeling assumptions regarding the material parameters and noise are stated explicitly \cite{stuart2010inverse,dashti2013bayesian}, providing a natural basis for the regularization of the problem, in contrast with ad-hoc regularization in classical approaches. Under assumptions of Gaussian prior and noise, which define the scope of this work, this problem is equivalent to a deterministic inverse problem with quadratic Tikhonov regularization. In a deterministic setting, however, there is no inherent basis for choosing these regularization terms. The Bayesian view does not remove these modeling choices, but it gives them a meaning in relation to statistics on the experimental noise and prior belief. Additionally, Bayesian treatment allows us to quantify the uncertainty in the inferred parameters through the posterior distribution. This can serve as the prior for a subsequent experiment, allowing iterative refinement of our modeling assumptions. Finally, and most importantly for this work, the Bayesian setting leads to  a natural objective for designing the next experiment, which is the expected information gain.

The availability of full-field measurements naturally raises the question of experimental design. Even though high-dimensional data are accessible, current experiments often rely on specimen geometries chosen using intuition or simple modifications of conventional geometries (e.g., dogbones and notched specimens) \cite{pierron2021towards}. These were originally developed for reliable testing and straightforward interpretation, but do not necessarily maximize the information content of the experiment. However, deliberately designing complex geometry specimens to probe multiple deformation states under a simple loading condition can provide richer information about the material parameters. This gives rise to an optimal experimental design problem-- the selection of a specimen geometry that maximizes information gain.  


A Bayesian framework provides a quantitative way to address optimal experimental design. A useful measure of experimental utility is the expected information gain (EIG) from the next experiment. It is defined through the expected Kullback--Leibler divergence between the prior and posterior distributions, provided by the Bayesian treatment \cite{lindley,chaloner1995bayesian}. Direct evaluation of the EIG is generally expensive because it involves averaging over all possible parameter values and future measurements. Gaussian or Laplace approximations are thus commonly used to obtain tractable local approximations \cite{long2013fast,wu2023fast}. Under the Gaussian and local assumptions considered here, the EIG criterion reduces to the classical D-optimal criterion, which minimizes the volume of the parameter uncertainty ellipsoid \cite{chaloner1995bayesian}. In the context of mechanical testing, D-optimal criteria have been used to optimize specimen geometries and loading paths for parameter identification \cite{bhattacharya2026optimal,ricciardi2024bayesian,chu2025bayesian}. Some approaches have also optimized other quantities such as parameter variance or deformation heterogeneity \cite{etling2018optimum,souto2016numerical,andrade2019design,bertin2016optimization,chapelier2022spline}. In much of this work, the design space is parametrized by a relatively small number of geometric variables, such as hole size, notch radius, or boundary shape. More recently, topology optimization has been explored for specimen design for material parameter identification, where the specimen geometry is represented by an element-wise density field. For linear elastic materials, various design criteria have been considered, including robustness of inversion, parameter sensitivity, and strain heterogeneity \cite{ghouli2025topology,chamoin2020coupling, barroqueiro2020design, gonccalves2023design}. The advantage of topology optimization is the large design space, since the material placement throughout the entire specimen can be varied \cite{bendsoe2013topology}, which enables non-intuitive specimen geometries.

In this paper, we propose a framework to design optimal specimen geometries with topology optimization for nonlinear material parameter identification. The main contribution of the present work is the integration of a Bayesian experimental-design criterion with density-based topology optimization for nonlinear constitutive parameter identification. The modeling assumptions (prior knowledge and experimental noise statistics) are specified explicitly and inform the design objective to maximize information for the next experiments. Since this design objective depends on the displacement solution and also on its sensitivity to the inferred parameters, computing design sensitivities is more involved than in a standard topology optimization practice. We therefore derive an adjoint-based formulation to efficiently compute the design sensitivities required to optimize the specimen. The framework accommodates full-field data, either displacements or raw images, together with force measurements.

We demonstrate the numerical framework using isotropic and anisotropic hyperelastic constitutive models under uniaxial tensile loading. The anisotropic example is a generic fiber-reinforced hyperelastic materials. These include soft biological materials, which exhibit strong anisotropy due to their underlying collagen fibers \cite{holzapfel2010constitutive, axer2001collagen}. For such materials, leveraging complex specimen geometry that excites heterogeneous deformation states is useful in identifying preferred directions and anisotropic material parameters. This approach requires full-field measurements (\cite{palanca2016use}) and cutting of the specimen into the optimized shape, which is possible for tissues such as skin \cite{annaidh2012characterization}. In general, fabrication and measurement constraints must be considered for each specific application.

The remainder of the paper is organized as follows. Section \ref{sec:framework_8} present the general abstract formulation, with the parameter identification formulated as a Bayesian inverse problem in Section \ref{sec:bip_8}, and optimal experimental design in Section \ref{sec:oed_8}.   We specialize to the problem of material parameter identification in Section \ref{sec:paramid_8}.  We introduce the general setting using topology optimization in \ref{sec:sett_8}, the material parameter identification in Section \ref{sec:param_8} and the optimal experimental design in Section \ref{sec:moed_8}.  The gradient based optimization approach, including the adjoint formulation, is developed in Section \ref{sec:adj_8}.  The implementation is summarized in Section \ref{sec:imp_8}.  The method is demonstrated in Section \ref{sec:demo_8} with isotropic materials in Section \ref{sec:iso_8} and anisotropic materials in Section \ref{sec:fiberimg_8}, and numerical studies in \ref{sec:numstudy}.

\section{General Framework} \label{sec:framework_8}
We provide the background for the problem of optimal design of experiments for material parameter identification in this section. We assume that we have a parametrized model, and therefore model discovery reduces to parameter identification.  We formulate this as a Bayesian inverse problem, and use it to formulate the problem of optimal experimental design.   This section draws from \cite{stuart2010inverse,lindley,alexanderian2016bayesian}.

\subsection{Parameter identification as a Bayesian inverse problem} \label{sec:bip_8}
We consider the Bayesian inverse problem of estimating unknown parameters $P\in\mathbb{R}^{d_P}$ from experimental data. For a given experimental design $\rho$, we assume that the measured data $\tilde{y}\in\mathbb{R}^d$ follows the model equation
\begin{equation} \label{eq:BIP_8}
    \tilde{y} = \mathcal{G}(P,\rho)+\eta,
\end{equation}
where $\mathcal{G}$ is parameter-to-observable map that maps the unknown parameter vector $P$ and the design variable $\rho$ to observations, and $\eta$ is the noise.  $\mathcal{G}$ involves the forward model of the experiment, the finite element discretization, and the observation operator, and is nonlinear in general \footnote{We provide a characterization of this map for our problem in Section \ref{sec:paramid_8}.}.  In this work, we assume that the noise is Gaussian with mean zero and covariance $\Gamma_{\text{noise}}\in\mathbb{R}^{d\times d}$, i.e., $\eta \sim \mathcal{N}(0,\Gamma_{\text{noise}})$.   Under these assumptions, the data likelihood, or the distribution of experimental data $\tilde{y}$ for a given parameter set $P$ and specified design $\rho$, is 
\begin{equation} \label{eq:likelihood_8}
    \pi_\text{like}(\tilde{y}|P;\rho) \propto  \exp\Biggr( -\frac{1}{2} (\mathcal{G}(P,\rho)-\tilde{y})^T\Gamma_{\text{noise}}^{-1}(\mathcal{G}(P,\rho)-\tilde{y})\Biggr).
\end{equation}

We assume that we have some prior knowledge about the unknown variables, typically obtained from previous experiments, existing measurements, or domain knowledge,   and then use $ \pi_\text{like}$ to refine our knowledge.  We assume that we have an uncorrelated Gaussian prior on the parameters, with measure $\mu_0 = \mathcal{N}(P_0,C_0)$, characterized by mean $P_0$ and covariance $C_0$. The prior mean represents the best guess about the unknown, while the covariance represents the uncertainty in that estimate. For the rest of this work, we only consider the case of a Gaussian prior and noise distribution.

According to Bayes' rule in our finite dimensional setting, the posterior distribution of the unknown parameter given the prior and the experimental observations admits a probability density \begin{equation}
    \pi_\text{post} (P|\tilde{y};\rho) \propto \pi_\text{like}(\tilde{y}|P;\rho)\,\pi_\text{prior}(P),
\end{equation}
where $\pi_\text{prior}$ is the probability density associated with prior measure $\mu_0$.   In general, this posterior probability distribution cannot be evaluated analytically when the forward map $\mathcal{G}$ is nonlinear.   Still, there are a couple of useful observations that we can make.

The value of the parameter that maximizes the posterior distribution, the maximum a posteriori (MAP) estimate, is given by the variational problem,
\begin{equation} \label{eq:objforp_8}
\begin{aligned}
    & P_\text{MAP} = \argmin \mathcal{O}_P(P), \\
    &\quad \text{ where }\mathcal{O}_P(P;\tilde{y},\rho) = \Phi(P;\tilde{y},\rho)   + \frac{1}{2} \Biggr( (P - P_0)^TC_0^{-1}(P - P_0) \Biggr), \,\\
    &\quad \quad \text{ and } \Phi(P;\tilde{y},\rho) = \Biggr( \frac{1}{2} (\mathcal{G}(P,\rho)-\tilde{y})^T\Gamma_\text{\text{noise}}^{-1}(\mathcal{G}(P,\rho)-\tilde{y})\Biggr).
 \end{aligned}
\end{equation}
The expression $\Phi$ in $\mathcal{O}_P$ is the negative log-likelihood (see Eq. \ref{eq:likelihood_8}) from a Bayesian perspective. From the perspective of classical optimization, this objective is the regularized data misfit functional $\Phi$ with a quadratic regularization.

Further, if $\mathcal{G}$ were linear, the posterior would be Gaussian with mean $P_\text{MAP}$.  Therefore, it is common to use a Laplace approximation (see \cite{long2013fast,wu2023fast,yoko2024optimal}) where we approximate the posterior with a Gaussian (or $\mathcal{G}$ with a linear map).  This Gaussian approximation of the posterior is $ \mathcal{N}(P_\text{MAP},C_\text{post})$ where 
\begin{equation} \label{eq:cpost_8}
    C_\text{post} = \bigg( H(P_{MAP}) + C_0^{-1} \bigg)^{-1},
\end{equation}
and $H$ is the Hessian of the negative log-likelihood ($\Phi(P;\tilde{y},\rho)$) evaluated at the MAP point.

\subsection{Optimal experimental design (OED)} \label{sec:oed_8}

The Bayesian inverse problem discussed previously assumes a fixed experimental design $\rho$.  We now seek to determine the experimental design $\rho$ such that the resulting observed data are maximally informative about the unknown variables.  We use the expected information gain (EIG) as the suitable utility function that provides us with a measure of the information content of the next experiment \cite{lindley}.  This expected information gain (EIG) is defined as the expected change in the Shannon entropy between the prior to the posterior distributions \cite{lindley} over all possible realizations of data. Equivalently, it can be defined by the expected Kullback-Leibler divergence (KL divergence) over all possible experimental data realizations\cite{alexanderian2016bayesian}. The KL divergence itself provides a distance measure between the posterior and prior probability density \cite{kullback1951information}.  We have,
\begin{align} 
    &EIG := \int_{\mathbb{R}^{d_P}}\int_{\mathbb{R}^d} D_{kl}(\pi_\text{post}(P|\tilde{y};\rho)||\pi_\text{prior})\pi_\text{like}(\tilde{y}|P;\rho) \, d\tilde{y} \,\pi_\text{prior}(P) \, dP, \\
    &\quad  \text{ where } D_{kl}(\pi_\text{post}||\pi_\text{prior}) = \int_{\mathbb{R}^{d_P}}\log \bigg( {\frac{\pi_\text{post}(P|\tilde{y};\rho)}{\pi_\text{prior}(P)} \bigg) \pi_\text{post} (P|\tilde{y};\rho) \, dP }.
\end{align} 

Note that the evaluation of the EIG requires knowledge of the posterior measure after obtaining data in the next experiment. Typically, this is computed by sampling across all possible future data realizations, which is numerically expensive. However, in the linear setting where $\mathcal G$ is linear with Gaussian noise and prior, the resulting posterior measure is also Gaussian with covariance given in (\ref{eq:cpost_8}). It follows that we can now characterize the EIG as
\begin{equation} \label{eq:DEIG_8}
    EIG \approx \mathcal{O}_\text{EIG} := \frac{1}{2}\log\bigg( \frac{\det C_0}{\det {C}_\text{post}}\bigg) = \frac{1}{2} \log \det(I + C_0 \, H(P_{MAP}) ).
\end{equation}
This is often commonly known as D-optimal criteria \cite{chaloner1995bayesian}, i.e., maximizing the EIG is equivalent to minimizing the determinant of the posterior covariance. We note that this expression is exact if $\mathcal{G}$ is linear; for $\mathcal G$ nonlinear, (\ref{eq:DEIG_8}) is obtained by linearizing about the current estimate and using the Gauss-Newton approximation for $H$.

As we shall see in the following section, $\mathcal{G}$ is nonlinear, and therefore, we adopt the local Bayesian D-optimal approximation. We start with the current best estimate of the parameter and the covariance; this is our prior $\mu_\text{prior}=\mathcal{N}(P_0,C_0)$. We then create a design to maximize the expected information gain  (\ref{eq:DEIG_8}), conduct an experiment, and then update our model parameters. This OED approach can be iterated, leading to a sequential local Bayesian D-optimal approach. After $k$ iterations, the prior parameter estimate $P_0 = P_k$ and prior covariance $C_0 = C_k$ act as priors in the problem to design specimen with $\rho_{k+1}$ for the next experiment.

\section{Application to materials parameter identification} \label{sec:paramid_8}

\subsection{Setting} \label{sec:sett_8}

We apply the framework of Section \ref{sec:framework_8} to the problem of identifying the material parameters of a hyperelastic material using mechanical tests.  We seek to design the shape of the specimen that enables us to optimally identify the material properties.  

We assume that our specimen is confined to a region $\Omega \subset {\mathbb R}^n$, which we regard as the maximal design region.  We specify the natural reference configuration of the body to be $\Omega_\chi = \{ x \in \Omega: \chi(x) = 1\}$ where $\chi: \Omega \to \{0,1\}$, and assume that the constitutive behavior of the body is described by a stored energy density $W(F;P)$ where $F = I+\nabla u$ is the deformation gradient, and $u$ is the displacement field.  We assume that the specimen is fixed at $\partial_0 \Omega$, and subject to a time-dependent Dirichlet boundary condition $y(x,t) = \bar{y}(x,t)$ on another part $\partial_u \Omega$ (so we restrict $\chi = 1$ on $\partial_u \Omega \cup \partial_0 \Omega$).   The remainder of the boundary of the specimen is traction-free.  We limit ourselves to the quasistatic setting and therefore the displacement $u:\Omega_\chi \to {\mathbb R}^n$ satisfies 
\begin{equation}  \label{eq:eq_8}
\nabla \cdot W_F (F; P) = 0 \text{ on } \Omega_\chi
\end{equation}
subject to the boundary conditions where $W_F = \partial W/\partial F$. 

Our goal is to design the specimen or the function $\chi$.  Unfortunately, optimal design problems are known to be nonconvex, and therefore we relax this problem using topology optimization and SIMP interpolation \cite{bendsoe2013topology} by introducing a SIMP exponent $p$.  We approximate $\chi$ using a function $\rho: \Omega \to  [\rho_\text{min}, 1]$ and assume that the body occupies the entire domain with energy density 
\begin{equation} \label{eq:SIMPW_8}
W(x,F;P) = \rho^p(x)  \, W(F;P).
\end{equation}
Equilibrium is given by (\ref{eq:eq_8}) on $\Omega$, or 
\begin{align} \label{eq:gov}
  -  \int_\Omega \left(\rho^p \ W_F (F; P) \cdot \nabla \varphi \right) \, d\Omega = 0  \quad \quad \forall \ \varphi \in \mathcal{U} = \{u \in H^1(\Omega), \,u = 0 \text{ on } \partial_0 \Omega \cup \partial_u \Omega \}.
\end{align}
Now, our design is given by the function $\rho$.

\subsection{Parameter identification} \label{sec:param_8}

We now specify the objective $\mathcal{O}_\text{P}$ that we use for parameter identification.  In a typical mechanical test, we decorate part of the boundary with a speckle pattern $g_0(x)$, and image the convected pattern $g(y,t)$ during deformation (where $y=x+u$), and also measure the reaction force $f(t)$ in some direction $e_f$ on $\partial_u \Omega$ where the displacement is applied.  There are two possible formulations.

\begin{itemize}

\item {\bf Image-based formulation} We follow Wihardja and Bhattacharya  \cite{wihardja2025constitutive} and consider the raw images and boundary forces as the experimental data $\tilde{y} = \{g(y,t), f(t)\}$.  We consider time-continuous measurements common in material testing (though this is easily specialized to a discrete time measurement).  The parameter to observable map (\ref{eq:BIP_8}) is
\begin{equation} \label{eq:forcing_8}
P \to \{ g(x+u(x,t),t), f_u(t)\} \quad \text{where } f_u (t) = \int_{\partial \Omega_u} W_F(\nabla u(x,t); P ) n\cdot e_f \, dA 
\end{equation}
is the reaction force on $\partial_u \Omega$ in the direction $e_f$ and $u$ solves the governing equation (\ref{eq:gov}).  Note that this map is nonlinear due to the nonlinearity of the governing equation (\ref{eq:gov}) as well as the speckle pattern $g$.  Therefore, the functional $\mathcal{O}_P$ in (\ref{eq:objforp_8} for Bayesian parameter identification becomes
\begin{equation}\label{eq:obja_8} 
\begin{aligned} 
    & \mathcal{O}_P = {\Phi}_{d} + {\Phi}_f + \frac{1}{2} (P - P_0)^T \quad C_0^{-1} (P - P_0),\\ 
    & \quad \Phi_d =  \frac{1}{2}\int_0^t \int_{\Omega } \rho^{2p} (g_0(x) - g(y(x,t),t))^T \quad \Gamma_i(t)^{-1} \, (g_0(x) - g(y(x,t),t)) \, d\Omega \, dt, \\ 
        & \quad\Phi_f =  \frac{1}{2}  \int_0^t \Bigg(f_p(t) - f_\text{exp}(t) \Bigg)^T \, \Gamma_f(t)^{-1} \Bigg(f_u(t) - f_\text{exp}(t) \Bigg) \ dt,
\end{aligned}
\end{equation}
where $\Phi = \Phi_d+\Phi_f$ is the total negative log-likelihood/data misfit with time-continuous data, $\Phi_d$ is the image mismatch between model prediction and experiments, $\Phi_f$ is the force mismatch, $\Gamma_i$ is the noise covariance associated with grayscale intensity values of images, and $\Gamma_f$ is the noise covariance associated with measured force. The prior mean of the parameter is $P_0$ with prior covariance $C_0$.  Note that we have weighted the images in $\Phi_d$ with $\rho$ since these are only observed in the actual design domain. 

\item {\bf Displacement-based formulation} It is common practice to use a two-step approach in experimental mechanics.  We first use the images to infer the displacement using Digital Image Correlation (DIC) \cite{sutton2009image,vic,yang_augmented_2020,hild}.  We then regard the displacement field as the experimental data in the parameter identification problem.   We note that the first inverse problem (image to displacement) is known to be ill-posed \cite{wihardja2025constitutive}, and so is regularized by constraints and filters.  Consequently, it is difficult to quantify the noise associated with the displacement field data.  In particular, the noise may be correlated based on the constraint.  Still, this is widely used, and therefore we study this as well.

In this setting, the data are $\tilde{y} = \{u_{DIC}(x,t), f(t)\}$ and the objective now becomes 
\begin{equation}\label{eq:objd_8} 
\begin{aligned} 
    & \mathcal{O}_P = {\Phi}_{d} + {\Phi}_f + \frac{1}{2} (P - P_0)^T \quad C_0^{-1} (P - P_0),\\ 
    & \quad \Phi_d =  \frac{1}{2} \int_0^t \int_{\Omega} \rho^{2p}(u_{DIC}(x,t) - u(x,t))^T \quad \Gamma_u(t)^{-1} \, (u_{DIC}(x,t) - u(x,t)) \, d\Omega \, dt,\\
        & \quad\Phi_f =  \frac{1}{2}  \int_0^t \Bigg(f_p(t) - f_\text{exp}(t) \Bigg)^T \, \Gamma_f(t)^{-1} \Bigg(f_u(t) - f_\text{exp}(t) \Bigg) \ dt,
\end{aligned}
\end{equation}
where $\Gamma_u$ is the noise covariance associated with the displacement-field measurements. 
\end{itemize}

In what follows, we use a displacement-based formulation for our development since this leads to more compact expressions.  However, the approach extends easily to the image-based formulation, and we evaluate both in our examples.

Given a proposed design $\rho$,  since no data from a $\rho$-experiment exists yet, we predict the covariance from that design at the current best estimate $P_0$,
\begin{equation} \label{eq:postfinal_8}
    C_\text{post}(\rho)   =\bigg( H(\rho) + C_0^{-1} \bigg)^{-1},
\end{equation}
where the Hessian in our setting is 
\begin{equation} \label{eq:hk_8}
\begin{aligned}
    H  & =  \int_0^t \Biggr( \int_{\Omega}  \left( \rho^{2p}  J ^T\Gamma_u^{-1} J \right) \, d\Omega +  J_f^T\Gamma_f^{-1}J_f  \Biggr) dt, \\
    & \text{with} \quad J = u_P := \frac{\partial u}{\partial P},\\
    & \phantom{\text{with}} \quad J_f = \frac{\partial f}{\partial P} = \int_{\partial_u \Omega}  \left(  W_{FF}(\nabla u, P_0) \nabla u_P + W_{FP}(\nabla u, P_0)  \right) \hat{n}  \cdot e_f   \ d A.
\end{aligned}
\end{equation} 
Once a design $\rho_\text{best}$ is selected and an experiment is performed, we can now compute the updated MAP estimate as the minimizer of (\ref{eq:objd_8}), 
\begin{equation} \label{eq:Pk1_8}
    P_\text{MAP} = \arg \min_P \  \mathcal{O_P}(P; u,\rho)  \quad \text{such that $u$ satisfies (\ref{eq:gov}) for $\rho$}. 
\end{equation}

Note that the Hessian $H$ is a functional of $\rho$, $u$ and $u_P$, or $H(\rho, u, u_P)$  at the current estimate $P_0$.  Also, (\ref{eq:Pk1_8}) is a PDE-constrained optimization problem, and (\ref{eq:hk_8}) requires the solution of (\ref{eq:gov}) as well as the sensitivity of this solution with respect to $P$.

\subsection{Optimal experimental design} \label{sec:moed_8}

We obtain the optimal experimental design for parameter identification under the local Bayesian D-optimal approximation.  At the start of the problem, we have the current parameter estimate $P_0$ and covariance $C_0$.  We can now specialize the expected information gain (\ref{eq:DEIG_8}) to our setting,
\begin{align} \label{eq:topopt_8}
  \rho_\text{best} = \argmin_{\rho} &-\mathcal{O}_{EIG}\bigg(\rho, u, u_P \,\bigg) \quad \text{such that $u$ satisfies (\ref{eq:gov}) for $\rho$, and} \\
  &\rho_\text{min} \leq \rho\leq 1, \quad\frac{1}{|\Omega|}\int_\Omega \rho(x) \,d\Omega = V_f \label{eq:designconstraint}.
\end{align}
Note that this is a PDE-constrained optimization problem where $u, u_P$ depend on $\rho$ through (\ref{eq:gov}). Additionally, our design constraint in (\ref{eq:designconstraint}) means the volume fraction $V_f$ remains the same in the initial and final designs.
\subsection{Gradient-based optimization with adjoint method} \label{sec:adj_8}

The considerations so far lead to a framework where we solve (\ref{eq:topopt_8}) for the design, perform an experiment, and then solve (\ref{eq:Pk1_8}) for refining current estimate of the parameters.  Note that both (\ref{eq:topopt_8}) and (\ref{eq:Pk1_8}) are PDE-constrained optimization problems.  We solve them using a gradient-based approach that exploits the adjoint method to compute the gradients. 

The functional $\mathcal{O}_\text{EIG}$ depends on $u$ and $\partial u/ \partial P$.  Therefore, we have two forward problems,
\begin{itemize}
\item[F1.] The displacement field $u$ satisfies
\begin{equation} \label{eq:res1_8}
\int_\Omega (\rho^p \ W_F(\nabla u, P_0)) \cdot \nabla\varphi \ d\Omega =0 \quad \forall \ \varphi \in \mathcal{U}.
\end{equation}
\item[F2.] The sensitivity of displacement to the parameter $u_P$ satisfies the following equation obtained by differentiating the above:
\begin{align} 
    \int_\Omega  \left(  \rho^p \left(W_{FP} (\nabla u, P_0)  +  W_{FF} (\nabla u, P_0) \nabla u_P \right) \right) \cdot \nabla \varphi   \, d\Omega  &= 0 \quad \forall \varphi \in \mathcal{U}, \label{eq:res2_8}
\end{align}
where $u$ is the solution of (\ref{eq:res1_8}).   We note in passing that (\ref{eq:res2_8}) may be viewed as an adjoint equation to (\ref{eq:res1_8}) if we were confined to parameter identification.
\end{itemize}

We now turn to finding the gradient of $\mathcal{O}_\text{EIG}$ with respect to $\rho$.  This requires care, since $u, u_P$ depend on $\rho$, and we use the adjoint method.  Given any $\lambda, \beta_P \in {\mathcal U}$, we use the forward problems (\ref{eq:res1_8} and (\ref{eq:res2_8}) to rewrite our objective as
\begin{align}
 \mathcal{O}^*(\rho,u, u_P)  &= {-\mathcal{O}_{EIG}}(\rho,u, u_P)  + \int_\Omega (\rho^p \ W_F(\nabla u, P_0))  \cdot \nabla \lambda\, d\Omega \\
    & \quad \quad + \int_\Omega  \left(  \rho^p \left(W_{FP} (\nabla u, P_0)  +  W_{FF} (\nabla u, P_0) \nabla u_P \right) \right)\cdot \nabla \beta_P \, d\Omega.
\end{align}
We perturb $\rho$ in the direction $\mu$ ($\rho \mapsto \rho+ \varepsilon \mu$ for small $\varepsilon$) and assume that the solutions of the forward problems F1 and  F2 are continuous with respect to $\rho$ so that the corresponding solutions are $u + \varepsilon v$ and $u_P + \varepsilon v_P$.  The variation of $\mathcal{O}^*$ with respect to $\rho$ in the direction $\mu$ is given by
\begin{equation}
\langle \delta_\rho \mathcal{O}^*, \mu \rangle := \left.\frac{d}{d\varepsilon} {\mathcal O}^*(\rho + \varepsilon \mu, u + \varepsilon v, u_P + \varepsilon v_P) \right|_{\varepsilon = 0}.
\end{equation}
A long, but straightforward calculation shown in Appendix \ref{app:adj_88} shows that the sensitivity of the objective to the design is given by
\begin{equation} \label{eq:oedsens_8}
\begin{aligned}
\langle \delta_\rho \mathcal{O}^*, \mu \rangle = & \int_\Omega   p \rho^{p-1} \bigg( -\frac{1}{2} C_0 (I+C_0H)^{-T}\cdot \int_0^t 2 \rho^p \,\,u_P^T \,\, \Gamma_u^{-1} \,\,u_P \, dt   \,\, \, \\
&\hspace{0.75in} + W_F(\nabla u, P_0)  \cdot \nabla \lambda  \\
&\hspace{0.75in} +  \Big(W_{FP} (\nabla u, P_0) +  W_{FF} (\nabla u, P_0) \nabla u_P \Big) \cdot \nabla \beta_P \bigg) \mu \, d\Omega,
\end{aligned}
\end{equation}
where $\lambda, \beta_P$ satisfy the following adjoint equations.
\begin{itemize}
\item[A1.]   The adjoint equation for $\beta_P$ is
\begin{equation} \label{eq:betaP_8}
\begin{aligned}
-\frac{1}{2}C_0 (I+C_0H)^{-T}\, \cdot \, \Biggr( &\int_0^t \int_{\Omega}  2 \rho^{2p} \,\, u_P \,\, \ \Gamma_u^{-1} \,\, v_P \,\, d \Omega \, dt\\
&+ \int_0^t 2\left(\int_{\partial_u \Omega} \left(  W_{FF} \nabla v_P \right)\hat{n} \cdot e_f\,  d A \, \right) \,\, \Gamma_f^{-1}\,\, J_f \, \, dt \Biggr)  \\
 &+ \int_\Omega  \left(  \rho^p \, \,W_{FF} \nabla v_P \right) \cdot \nabla \beta_P \, d\Omega = 0  \quad  \forall \ v_P \in \mathcal{U}.
\end{aligned}
\end{equation}
\item[A2.] With $\beta_P$ as in (\ref{eq:betaP_8}), the adjoint equation for $\lambda$ is
\begin{equation} \label{eq:lambdaadj_8}
\begin{aligned} 
&-\frac{1}{2}C_0(I+C_0H)^{-T} \, \cdot \, \Biggr( \int_0^t 2 \,  J_f^T \,\, \Gamma_f^{-1} \,   \int_{\partial_u \Omega} \left( \left( W_{FFF} \nabla u_P + W_{FFP} \right) \nabla v \right)\hat{n}\cdot e_f \,\, dA \,\,\, dt \Biggr)\\
& \quad + \int_\Omega \rho^p \left( W_{FF}\nabla v  \right) \cdot  \nabla \lambda\\
& \quad + \int_\Omega \rho^p  \left( W_{FFF}\nabla v \nabla u_P  + W_{FFP}\nabla v \right)\cdot \nabla \beta_P  \,\, d\Omega = 0 \quad  \forall \ v \in \mathcal{U}.
\end{aligned}
\end{equation}
\end{itemize}

\subsection{Implementation} \label{sec:imp_8}

We choose a suitable reference domain or ground structure that allows for the definition of boundary conditions. We specify parts of the ground structure that can be designed or left as solids/voids. We construct a sufficiently fine finite element mesh of the ground structure such that the structure can be described in a reasonable bit-map representation. This mesh is unchanged throughout the entire optimization procedure. We discretize the displacement finite element as a standard piecewise polynomial Lagrange basis function of degree 1, and the design variable $\rho$ as piecewise constant on each element.   All PDEs are solved using a finite element method in deal.II \cite{dealII94}.  

In performing topology optimization, we use the SIMP interpolation method \cite{bendsoe1999material}, where the design density is scaled by an interpolation power $\rho^p$. We initialize the optimization with $p=1$ and increase it gradually to $p_{\max}=3$, following the standard practice \cite{bendsoe2013topology} to promote convergence toward a solid--void design. In all examples, $p$ is increased by $0.1$ every 20 iterations. 
At each optimization iteration, we use the method of moving asymptotes (MMA) \cite{svanberg1987method} based on the sensitivity (\ref{eq:oedsens_8}) to update the design.   We then filter it with a renormalized filter with a linear weight function \cite{bourdin2001filters} to avoid the development of fine-scale structures due to the ill-posedness of the topology optimization. Once $p_{\max}$ is reached, the optimization is continued until either the design sensitivity or the incremental objective reaches $\varepsilon = 10^{-4}$ and the design is sufficiently well-defined. We summarize our approach in  Algorithm \ref{alg:algo_8}. The filtering method used here introduces an effective regularization length, which can be adjusted depending on knowledge of fabrication constraints or limitations. Throughout this work, the filter radius is chosen to be $4\times h$, where $h$ is the element mesh size.

All computations were performed on a workstation with a 12th Gen Intel Core i9-12900 CPU (16 cores / 24 threads, up to 5.1 GHz) and 64 GB RAM. For the mesh size used here (96 $\times$ 32 elements, with a total of 6402 DOFs), one MMA iteration consisting of the two forward solves $(F1,F2)$, assembly of the Hessian, two adjoint solves $(A1,A2)$, and the design update itself requires approximately 50 seconds. Convergence of the final solid-void design (SIMP exponent increased from 1 to 3 over 400 iterations) therefore took approximately 5 hours on this hardware.

\begin{algorithm}[t]
\caption{Optimal Experimental Design \label{alg:algo_8}}
\textbf{Input:} Reference domain $\Omega$ with triangulation, current estimate of parameters of the model $P_0$ with covariance $C_0$ and current design $\rho$  \\
\textbf{Output:} Updated design field $\rho_\text{best}$
\begin{algorithmic}
    \State \textbf{Initialize} Design $\rho(x)$, current model parameters $P_0$, and initial SIMP power $p=1$;
    \While{$p < p_{\max}$ \textbf{and} $\| \delta_\rho \mathcal{O}_\text{EIG} \| \, \textbf{or} \, \|\mathcal{O}_{\text{EIG},k+1}-\mathcal{O}_{\text{EIG},k1}\| > \varepsilon$}
        \State Step 1: Filter the design density field $\rho$;
        \State Step 2: Solve forward problems F1, F2 for  $u$ and  $u_P$;
        \State Step 3: Compute $C_\text{post}(\rho)$ (Eq. \ref{eq:postfinal_8}) and objective $\mathcal{O}_{\text{EIG}}$ (Eq. \ref{eq:DEIG_8}); 
        \State Step 4: Solve the adjoint problems  A1, A2 for $\beta_P$ and $\lambda$;
         \State Step 5: Compute design sensitivity    $\delta_\rho\mathcal{O}_{\text{EIG}}$ at each element;
        \State Step 6: Update design density field $\rho$ with MMA at each element;
        \State Step 7: Every 20 MMA iterations, update SIMP power $p \leftarrow \min(p+0.1,p_{\max})$
    \EndWhile
\end{algorithmic}
\end{algorithm}

\section{Demonstration on hyperelastic materials} \label{sec:demo_8}

\begin{figure}
\centering
\includegraphics[width=5.5in]{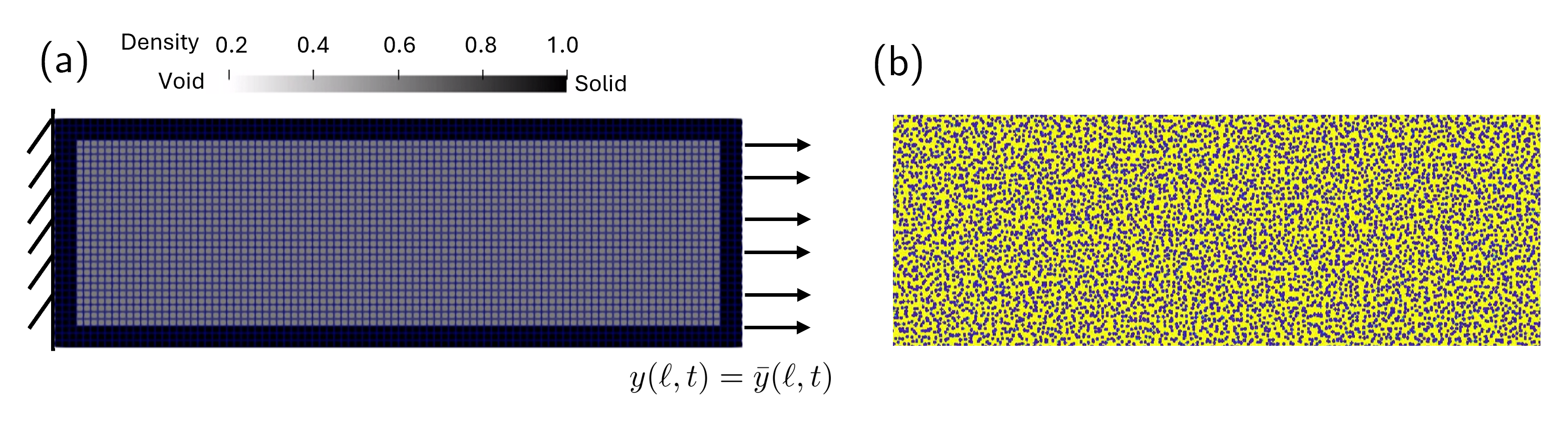}
\caption{ (a) Reference domain of $0.036\times0.012$ m$^2$, with finite element mesh consisting of $96\times32$ elements. An example density assignment to initiate the optimization can be seen in the scale bar. (b) Representative of speckle pattern to perform image-based optimal designs ($15\times15$ pixels within each mesh), with $4\times4$ pixels per speckle.} 
\label{fig:oedsetup_8}
\end{figure}

We now demonstrate the proposed approach on hyperelastic materials.  We demonstrate both the image-based and displacement-based approaches.  We study two examples, one isotropic and one anisotropic.  In all our examples, the reference domain is rectangular, $\Omega=0.036\times0.012$ m$^2$, with one edge held fixed and the other subjected to uniaxial extension as shown in Figure \ref{fig:oedsetup_8}.  In some of our examples, we fix a strip near the boundary of the domain to be fixed at $\rho=1$ (solid), as shown in Figure \ref{fig:oedsetup_8}(a), and design only the interior.  In other examples, we only fix the edges at the grips.

We demonstrate a single step of the procedure of optimal design and show that this already leads to meaningful designs. As mentioned earlier, one can perform the proposed optimal design framework iteratively, alternating between designing optimal specimens and refining parameter estimates.

\subsection{Isotropic material} \label{sec:iso_8}
We assume
\begin{equation}
     W = \frac{\mu}{2}  \ \text{tr}\big(F^TF - 3 - 2\log J \big) + \frac{k}{2} \ ( J-1 )^2,
\end{equation}
where $J =\det F$ in three dimensions, and specialize to plane stress.  Under plane stress, the deformation gradient is in plane, with the only nonzero out-of-plane stretch being $F_{33}$, determined locally from $\sigma_{33} = 0$. The parameters to be estimated are $P= \{ \mu,k\}$. 

We assume that our prior (assumed Gaussian) are characterized by
\begin{equation}
P_0=\{ \mu^0,k^0\}=\{4 \,\text{MPa}, 8.67 \,\text{MPa} \}, \quad C_0 = \begin{bmatrix}
    \sigma_\mu^2 & 0 \\
    0 & \sigma_k^2
\end{bmatrix}, \ \  \sigma_\mu = \sigma_k = 1 \,\text{MPa}.
\end{equation}
Similarly, we assume that our noise (assumed Gaussian and uncorrelated) has zero mean with covariance
\begin{equation}
\Gamma_i= 0.01^2\,  \mathbf{I}, \quad \Gamma_u = (10^{-5})^2\,  \mathbf{I}, \quad \Gamma_f= (0.01 t)^2\,  \mathbf{I}\, 
\end{equation}
for the images (in image-based formulation), displacements (in displacement-based formulation) and force, respectively.  Note that we have taken the force to increase with time; we anticipate our force to increase, and the noise on the typical load cells increases with force magnitude. Note that we consider a uniaxial extension setting, so we take $e_f = e_1$ in \ref{eq:forcing_8} and neglect the contribution of the other components. 

\begin{figure}
\centering
\includegraphics[width=6.0 in]{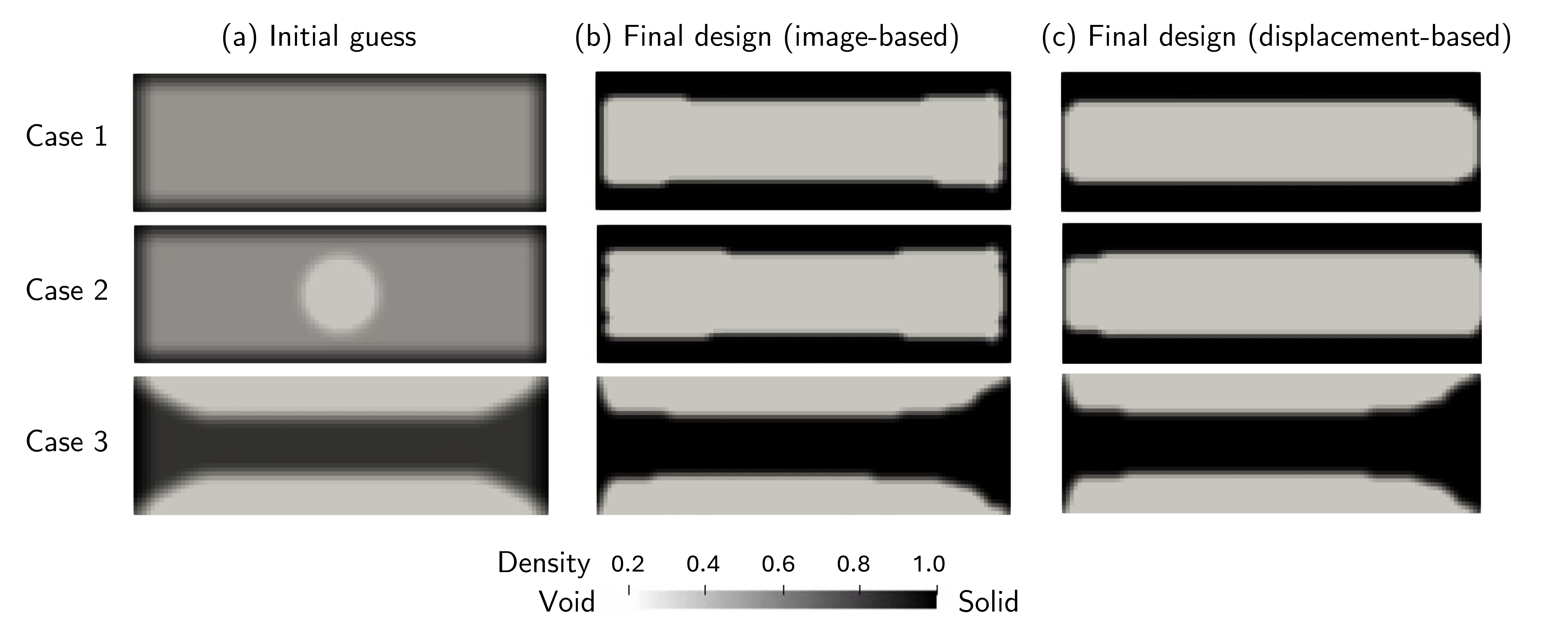}
\caption{Optimal design for isotropic material. (a) Initial design (density assignment) for optimization. (b) Resulting optimal topology with images as raw data, and (c) with displacements as raw data.}
\label{fig:isotropic_8}
\end{figure}

The results of our simulations are shown in Figure \ref{fig:isotropic_8}.   The top row shows the designs when we constrain the entire boundary to have fixed density and start with a uniform initial guess.  The second row retains the same constraint, but starts with an initial configuration consisting of a hole.  The resulting designs are similar, showing that in this case the design is independent of the initial guess.  Both the image- and displacement-based formulations lead to initial designs that have two strips that are largely uniform through the length.  
The third row shows a design when we use an initial dogbone design and remove the constraint on the top and bottom.

We follow the SIMP interpolation as described in Section \ref{sec:imp_8} and increase the SIMP exponent from 1 to 3 over the first 400 iterations. We obtain the final solid-void designs within the next 100 to 400 iterations, depending on the initial design guess.

We now turn to evaluate the performance of the optimal design in inferring parameters from experiments by using synthetic data. We set the reference parameters and perform numerical simulations to compute the experimental observables (image and force) and add noise to them consistent with our assumptions. Then, these synthetic data are used to recover the parameters following the image-based inference procedure \cite{wihardja2025constitutive}. We chose the image-based design (Case 1 of Figure \ref{fig:isotropic_8}b) as an example to compare with a random design with the same volume fraction. The inferred shear modulus has a 0.3\% error for both the optimal design and the random design, while the error in bulk modulus is 4\% and 10\% for the optimal and random design. We see that the optimized design can lead to more accurate parameter inference.


\subsection{Fiber-reinforced material} \label{sec:fiberimg_8}

We next consider an anisotropic hyperelastic material with two fiber families in the plane of the specimen, with stored energy density
\begin{align}
     W =   C_1(\bar{I_1}-3) + C_4(\bar{I_4}-1)^2 + C_6(\bar{I_6}-1)^2 + \frac{\kappa}{2}(J-1)^2 ,
\end{align}
where $\bar{I_1} =  \text{tr}(\bar{C}), I_4 = a \cdot \bar{C} a, I_6 = b \cdot \bar{C} b$, $J= \det F$ for $\bar{C} = J^{-4/3}C$, $a = (\cos \theta_1, \sin \theta_2, 0)$ and $b = (\cos \theta_2, \sin \theta_2,0)$.  We specialize to plane stress.  The parameters to be estimated are $P= \{ C_1,C_4,C_6,\kappa,\theta_1,\theta_2\}$.  

Unless otherwise stated, the mean and covariance of our prior (Gaussian and uncorrelated) are given by
\begin{equation} \label{eq:base_8}
\begin{aligned}
&P_0=\{ C_1^0, C_4^0,C_6^0,\kappa^0,\theta_1^0,\theta_2^0\} = \{4 \text{ MPa},8 \text{ MPa},8 \text{ MPa},13 \text{ MPa},\pi/4,-\pi/6\}, \\
&C_0= \text{diag}(\sigma_{C_1}^2, \sigma_{C_4}^2 ,\sigma_{C_6}^2 ,  \sigma_{\kappa}^2, \sigma_{\theta_1}^2, \sigma_{\theta_2}^2),
\end{aligned}
\end{equation}
where the variance is $20\%$ of mean value (for example, $\sigma_{C_1} = 20 \, \%C_1^0 = 0.8 \text{MPa}$)  . We assume a Gaussian noise distribution $\eta  \sim \mathcal{N}(0,\Gamma)$ for all data measurements.

The image-based formulation often leads to non-convergence. The image-matching contribution to \eqref{eq:hk_8} is
\begin{equation}
\int_0^T \left( \int_\Omega \rho^{2p} \left(\frac{\partial g}{\partial y}\frac{\partial u}{\partial P}\right)^T \Gamma_i^{-1} \left(\frac{\partial g}{\partial y}\frac{\partial u}{\partial P}\right) d\Omega + J_f^T \Gamma_f^{-1} J_f \right) dt,
\end{equation}
and thus the design sensitivity requires us to differentiate the images, $\partial g/\partial y$. A speckle pattern varies rapidly at the pixel scale, so this gradient is noisy and high-frequency, and the resulting element-wise sensitivity is highly oscillatory. Therefore, we confine ourselves to the displacement-based formulation in this section.

 We assume spatio-temporally uncorrelated Gaussian noise  $\Gamma_f= (0.01 t)^2\,  \mathbf{I}$ for the force measurements: as before, it increases with time since the noise in a typical load cell increases with force.  We consider two types of noise for the displacement, both spatio-temporally uncorrelated and Gaussian.   The first is constant noise $\Gamma_u = (10^{-5})^2\,  \mathbf{I}$, and the second is linearly increasing $\Gamma_u = (10^{-6}\times t)^2\,  \mathbf{I}$ to reflect the increasing noise with increasing displacement in a typical DIC inversion.

\begin{table}
\centering
\small
\begin{tabularx}{\textwidth}{cX}
\toprule
Case & Description \\
\midrule
\multicolumn{2}{c}{\textbf{Various priors}}\\
\addlinespace
P1 &
Baseline (\ref{eq:base_8}): $\{C_1^0,C_4^0,C_6^0,\kappa^0,\theta_1^0,\theta_2^0\}
=\{4~\mathrm{MPa},\,8~\mathrm{MPa},\,8~\mathrm{MPa},\,13~\mathrm{MPa},\,\pi/4,\,-\pi/6\}$ \\
P2 &
$\theta_1^0=\pi/3$, remaining parameters identical to the baseline. \\
P3 &
$C_6^0=16~\mathrm{MPa}$, remaining parameters identical to the baseline. \\
P4 &
$\{C_1^0,C_4^0,C_6^0,\kappa^0\}
=\{4.5~\mathrm{MPa},\,9~\mathrm{MPa},\,9~\mathrm{MPa},\,15~\mathrm{MPa}\}$,
remaining parameters identical to the baseline. \\
\midrule
\multicolumn{2}{c}{\textbf{Various initial designs and constraints}}\\
\addlinespace
C1 & Uniform initial design; edge constraint. Volume fraction, $V_f = 0.6.$ \\
C2 & Initial design I1; edge constraint.  Volume fraction, $V_f = 0.6$.\\
C3 & Initial design I2; edge constraint.  Volume fraction, $V_f = 0.6$.\\
C4 & Initial dogbone design; grip constraint.  Volume fraction, $V_f = 0.5$.\\
\bottomrule
\end{tabularx}
\caption{Cases considered using the displacement formulation.}
\label{tab:cases}
\end{table}

\begin{figure}
    \centering
     \includegraphics[width=4.0in]{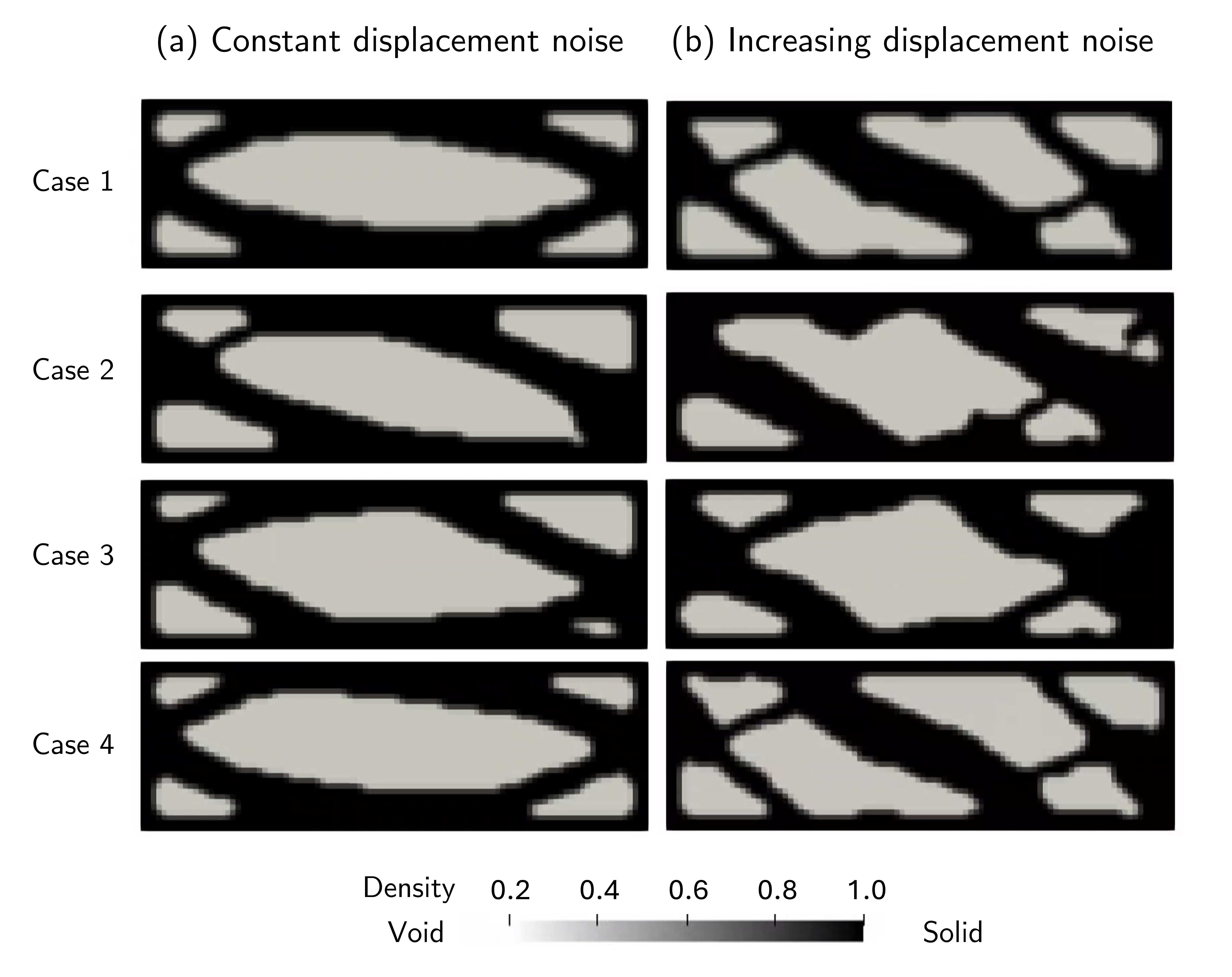}
    \caption{Optimal design for fiber-reinforced hyperelastic model in the displacement formulation for different priors.  (a) constant and (b) increasing noise.  The rows depict different priors as noted in Table \ref{tab:cases}. All these designs use C1: uniform density  as an initial guess and constrain the edges to be solid.}
    \label{fig:des_priors}
\end{figure}

\begin{figure}
    \centering
     \includegraphics[width=6.0in]{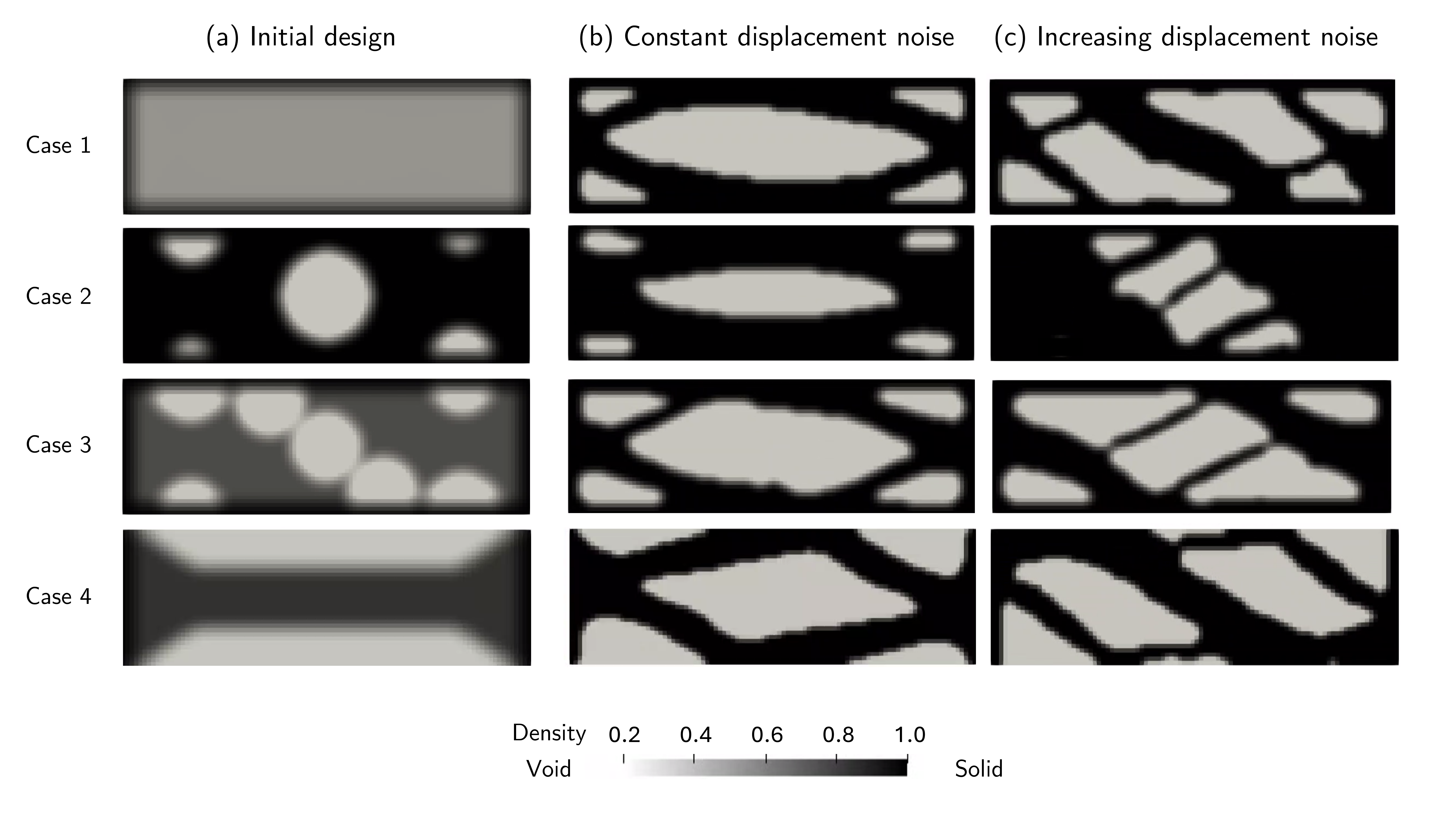}
    \caption{Optimal design for fiber-reinforced hyperelastic model in the displacement formulation of different initial designs and constraints.
    (a) initial design,  (b) constant and (c) increasing noise. The rows depict different initial designs and constraints as noted in Table \ref{tab:cases}. All these designs use P1 as prior.}
    \label{fig:des_init}
\end{figure}

The results are shown in Figure \ref{fig:des_priors} for various priors as noted in Table \ref{tab:cases}.   Each of these designs is subject to the constraint that the edges are solid, and the initial guess is uniform.  We observe that despite the fact that the priors vary by about 15\%, the resulting designs are similar.  This is significant since it suggests that we get meaningful designs even without iteration.  Turning to the designs, we observe that they all have linear elements aligned in multiple directions to be able to assess the anisotropy.  Finally, the uniform displacement noise results in a simpler design compared to the case of increasing displacement noise.

Figure \ref{fig:des_init} shows the result for various initial designs and for different design constraints, as noted in Table \ref{tab:cases}. 
As expected of a typical nonconvex topology optimization problem, the resulting final designs depend on the initial starting designs. The resulting solutions should thus be interpreted as local optima instead of globally optimal designs. This is reasonable within the scope of our framework, which is a local experimental design method to identify an informative next specimen. Still, while the details of the design depend on these initial guesses and constraints, the overall designs are all very similar in that they all have linear elements aligned in multiple directions to be able to assess the anisotropy. In addition, the expected information gain, $\mathcal{O}_\text{EIG}$, is similar across all cases with the same volume fraction. The $\mathcal{O}_\text{EIG}$ for designs C1, C2, C3 are 12, 12, 11.9 respectively for constant displacement noise and 11, 10.7, 10.7 for increasing displacement noise.

We now turn to understanding the efficacy of our optimal designs.  As a representative example, we consider the optimal grip-constrained design (Design C4, Figure \ref{fig:des_init}).  For constant noise, the expected information gain $\mathcal{O}_\text{EIG}$ increases from 9.5 for the initial to 11.3 for the final.  For increasing noise, $\mathcal{O}_\text{EIG}$ increases from 11 for the initial to 12 for the final.  Thus, the optimization does lead to an increase in the expected information gain.  We compare the performance of our optimized design with a number of arbitrary designs in Table \ref{table:sum_88}.  In each case, the optimized design has the highest expected information gain.

Another indicator is the trace of the expected posterior covariance.  This is often termed the A-optimal criterion that minimizes the mean square error of the inferred parameters \cite{ucinski2004optimal}. We split this into the trace of the part related to moduli and the part related to orientations, and note that these are related to mean square error. These are also shown in Figure \ref{table:sum_88}.  Again, in each case, we notice that our optimized designs perform better than the arbitrary design.  Thus, we observe that optimizing under the D-optimal criterion also leads to a better performance on the A-optimal criterion.
 
These indicators above depend only on the prior and do not involve any experimental evaluation.  However, our goal is to infer the parameters from the experiment.  We use synthetic data to evaluate this.  We use a set of reference parameters and numerical simulation to compute the experimental observables (image, displacement and force), and add noise to it consistent with our assumptions to create synthetic data.  We use this synthetic data to recover our parameters.  We can do so in two ways: either using the image-based formulation or the displacement-based formulation for parameter inference\footnote{The designs are optimized using displacement-based formulation for both cases.}.  Table \ref{table:inf_88} shows the error in the inferred parameters,
\begin{equation}
\frac{P_\text{MAP} - P_\text{true}}{P_\text{true}},
\end{equation}
 for the optimal design (Design C4 in Figure \ref{fig:des_init}) for two sets of reference parameters.   Finally, this table also compares the inference using a couple of arbitrary designs (Design A2 and A4 in Table \ref{table:sum_88}) with that of the optimized design in Design C1 (Figure \ref{fig:des_init}) under constant noise.    We see that the optimized designs lead to more accurate parameter inference in each case. 
 
 In summary, our optimization leads to better expected information gain as well as more accurate parameter inference.


\begin{table}
\centering
\small
\begin{tabularx}{\textwidth}{p{2.0cm}*{7}{>{\centering\arraybackslash}X}}
\toprule
\multicolumn{7}{c}{\textbf{Edge constraint}, Volume fraction $V_f=0.6$} \\
\midrule
 & Design C1 & Design A1 & Design A2 & Design A3 & Design A4 & Design A5 \\
\midrule
\multicolumn{7}{l}{\emph{Constant noise}} \\
$\mathcal{O}_{EIG}$ & 12 & 10.5 & 11.4 & 10.4 & 11 & 10.6 \\
MSE$_{\text{mod}}$ & 8.2 & 10 & 9.75 & 10 & 9.6 & 10 \\
MSE$_{\text{orient}}$ & 0.002 & 0.009 & 0.006 & 0.012 & 0.006 & 0.01 \\
\addlinespace
\multicolumn{7}{l}{\emph{Increasing noise}} \\
$\mathcal{O}_{EIG}$ & 11 & 9.7 & 10 & 9.6 & 9.9 & 9.8 \\
MSE$_{\text{mod}}$ & 10 & 10.8 & 10.7 & 10.9 & 10.9 & 11 \\
MSE$_{\text{orient}}$ & 0.005 & 0.04 & 0.035 & 0.035 & 0.034 & 0.034 \\
\midrule

\multicolumn{7}{c}{\textbf{Grip constraint}, Volume fraction $V_f=0.5$} \\
\midrule
 & Design C4 & Design B1 & Design B2 & Design B3 & Design B4 & Design B5 \\
\midrule
\multicolumn{7}{l}{\emph{Constant noise}} \\
$\mathcal{O}_{EIG}$ & 12 & 11 & 10.9 & 9.9 & 10.4 & 11 \\
MSE$_{\text{mod}}$ & 8.7 & 10 & 10.2 & 10.4 & 9.8 & 9.8 \\
MSE$_{\text{orient}}$ & 0.003 & 0.008 & 0.0083 & 0.016 & 0.01 & 0.007 \\
\addlinespace
\multicolumn{7}{l}{\emph{Increasing noise}} \\
$\mathcal{O}_{EIG}$ & 11.3 & 9.6 & 9.6 & 9.4 & 9.7 & 9.9 \\
MSE$_{\text{mod}}$ & 9.8 & 11 & 11 & 11 & 11 & 10.8 \\
MSE$_{\text{orient}}$ & 0.001 & 0.06 & 0.07 & 0.04 & 0.025 & 0.04 \\
\midrule

\multicolumn{7}{l}{\textbf{Glossary:}}\\
\multicolumn{7}{l}{MSE$_{\text{mod}}=$ Posterior $(\sigma_{C_1}^2+\sigma_{C_4}^2+\sigma_{C_6}^2+\sigma_\kappa^2)$ MPa$^2$}\\
\multicolumn{7}{l}{MSE$_{\text{orient}}=$ Posterior $(\sigma_{\theta_1}^2+\sigma_{\theta_2}^2)$}\\[0.5em]
\multicolumn{7}{c}{\includegraphics[width=6.0in]{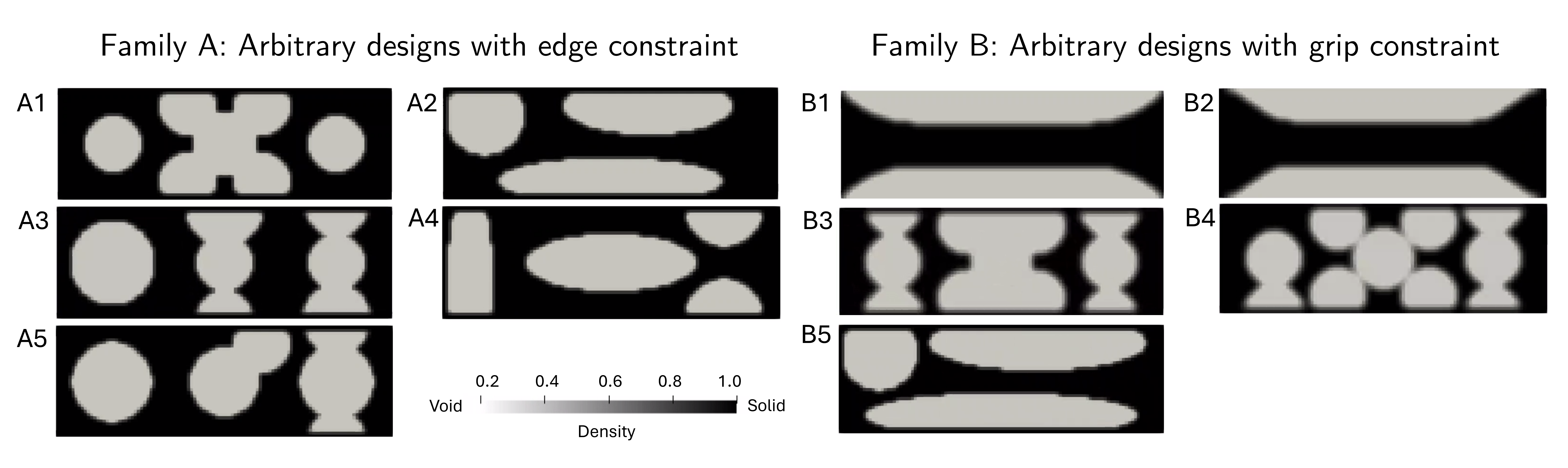}}\\
\bottomrule
\end{tabularx}

\caption{Evaluation of the optimal designs: comparison of the expected information gain and uncertainty reduction of the optimized designs against those of arbitrary designs.  All cases use prior P1.}
\label{table:sum_88}
\end{table}


\begin{table}
\centering
\small
\begin{tabularx}{\textwidth}{p{6.0cm}*{6}{>{\centering\arraybackslash}X}}
\toprule
& $C_{1,\text{err}}(\%)$ & $C_{2,\text{err}}(\%)$ & $C_{3,\text{err}}(\%)$ & $C_{4,\text{err}}(\%)$ & $\theta_{1,\text{err}}(\%)$ & $\theta_{2,\text{err}}(\%)$ \\
\midrule
\multicolumn{7}{c}{\textbf{Design C4: Displacement-based design, Displacement-based recovery}}\\
\addlinespace
\multicolumn{7}{l}{%
\textbf{$P_{\text{true}}$}=$\{C_1=4.5,\ C_2=C_3=9,\ C_4=14.6~\mathrm{MPa},\ \theta_1=40^\circ,\ \theta_2=27^\circ\}$}\\
\addlinespace
Initial dogbone & 3.1 & 2.2 & 2.6 & 2.6 & 15 & 0.3\\
Final (constant noise) & 2.6 & 1.0 & 1.6 & 1.2 & 15 & 0.3\\
\addlinespace
\multicolumn{7}{l}{%
\textbf{$P_{\text{true}}$}=$\{C_1=3.5,\ C_2=C_3=7,\ C_4=11.4~\mathrm{MPa},\ \theta_1=43^\circ,\ \theta_2=31^\circ\}$}\\
\addlinespace
Initial dogbone & 3.7 & 6.7 & 5.9 & 6.4 & 9.6 & 3.8\\
Final (constant noise) & 0.6 & 5.5 & 4.6 & 5.5 & 7.4 & 3.7\\
\midrule

\multicolumn{7}{c}{\textbf{Design C4: Displacement-based design, Image-based recovery}}\\
\addlinespace
\multicolumn{7}{l}{%
\textbf{$P_{\text{true}}$}=$\{C_1=4.5,\ C_2=C_3=9,\ C_4=14.6~\mathrm{MPa},\ \theta_1=40^\circ,\ \theta_2=27^\circ\}$}\\
\addlinespace
Initial dogbone& 2.7 & 3.7 & 9.8 & 3.0 & 4.0 & 8.8\\
Final (constant noise) & 2.2 & 2.3 & 3.5 & 1.5 & 4.0 & 8.7\\
Final (increasing noise) & 2.2 & 0.3 & 1.8 & 2.0 & 5.0 & 8.75\\
\addlinespace
\multicolumn{7}{l}{%
\textbf{$P_{\text{true}}$}=$\{C_1=3.5,\ C_2=C_3=7,\ C_4=11.4~\mathrm{MPa},\ \theta_1=43^\circ,\ \theta_2=31^\circ\}$}\\
\addlinespace
Initial dogbone& 5.0 & 13.6 & 5.0 & 4.8 & 8.9 & 4.3\\
Final (constant noise) & 0.7 & 6.6 & 6.6 & 4.0 & 4.7 & 4.2\\
Final (increasing noise) & 4.0 & 6.6 & 5.0 & 3.4 & 2.4 & 4.2\\
\midrule

\multicolumn{7}{c}{\textbf{Design C1 vs. arbitrary: Displacement-based design, Image-based recovery}}\\
\addlinespace
\multicolumn{7}{l}{%
\textbf{$P_{\text{true}}$}=$\{C_1=4.5, C_2=C_3=9, C_4=14.6\ \text{MPa} ,\ \theta_1=40^\circ,\ \theta_2=27^\circ\}$} \\
\addlinespace
Design A2 (Table~\ref{table:sum_88}) & 4.4 & 7.6 & 8.4 & 3.0 & 6.5 & 8.8\\
Design A4 (Table~\ref{table:sum_88}) & 5.4 & 6.4 & 6.2 & 2.3 & 6.0 & 8.8\\
Optimal design, Design C1 (Fig. \ref{fig:des_priors}) & 4.4 & 5.9 & 6.1 & 1.9 & 3.8 & 8.8\\
\bottomrule
\end{tabularx}
\caption{Evaluation of the optimal designs: comparison of the percentage error in parameter inference using the optimal designs versus non-optimal designs.}
\label{table:inf_88}
\end{table}

\subsection{Numerical study} \label{sec:numstudy}
\begin{figure}
  \centering
  \includegraphics[width=0.9\textwidth]{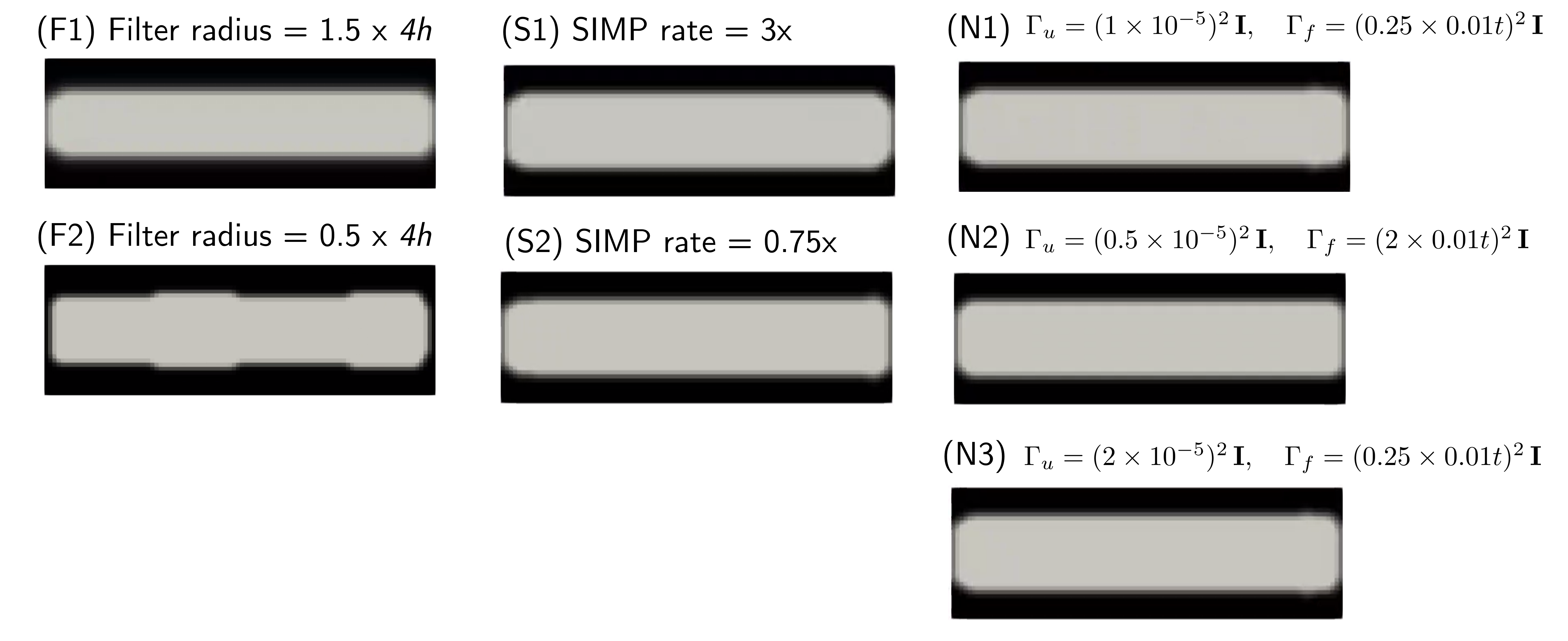}
  \caption{Optimal design for the isotropic hyperelastic material under different filter radii (F1, F2), different SIMP exponent increase rates (S1, S2), and different noise magnitudes (N1, N2). The baseline optimal design is obtained with $4h$ filter radius, SIMP exponent increase of $0.1$ every $20$ iterations, and noise level $\Gamma_u = (10^{-5})^2\mathbf{I}$, $\Gamma_f = (0.01t)^2\mathbf{I}$. }
  \label{fig:numstudy_iso}
\end{figure}

\begin{figure}
  \centering
  \includegraphics[width=0.9\textwidth]{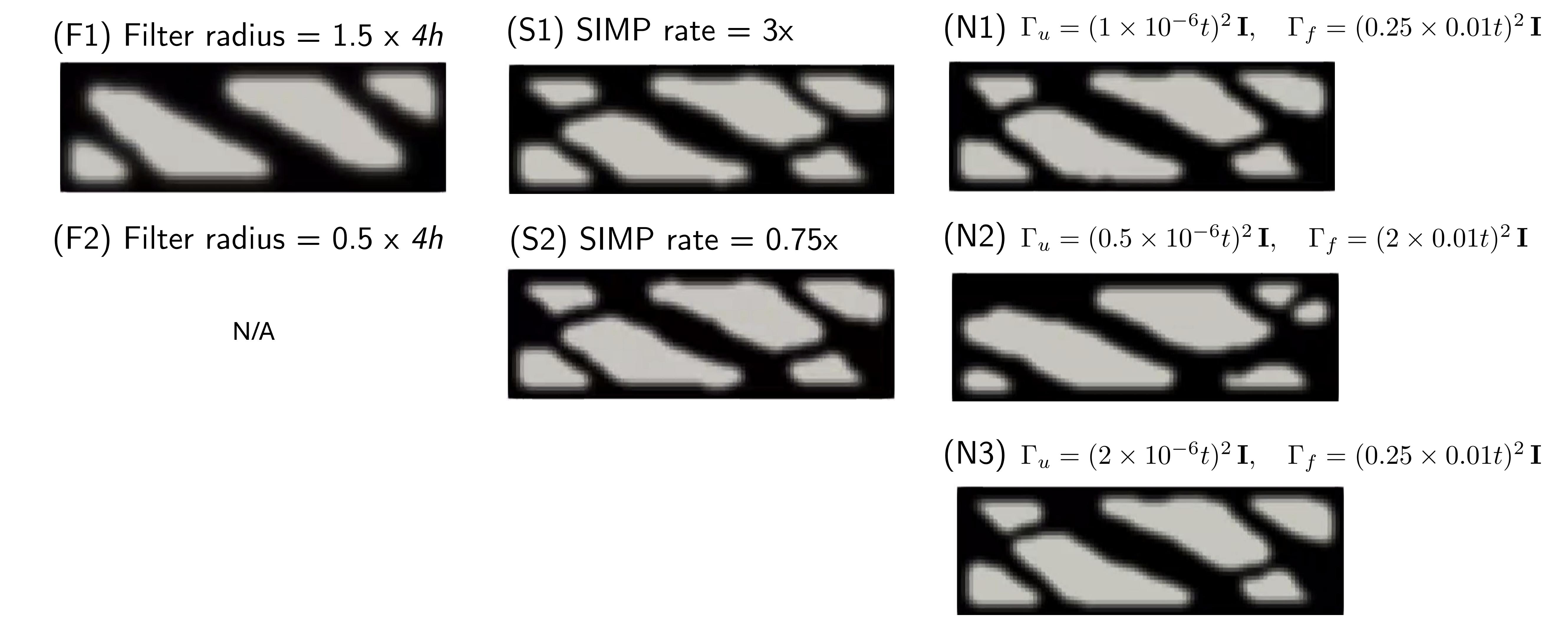}
  \caption{Optimal design for the anisotropic hyperelastic material under different filter radii (F1, F2), different SIMP exponent increase rates (S1, S2), and different noise magnitudes (N1, N2). The baseline optimal design is obtained with $4h$ filter radius, SIMP exponent increase of $0.1$ every $20$ iterations, and noise level $\Gamma_u = (10^{-6}t)^2\mathbf{I}$, $\Gamma_f = (0.01t)^2\mathbf{I}$. }
  \label{fig:numstudy_aniso}
\end{figure}

We now examine how our optimal designs changes with respect to numerical parameters of the topology optimization scheme and to the assumed noise levels. We consider the displacement-based formulation and all cases start from a uniform initial guess design (C1) and prior P1 in Table \ref{tab:cases}. The baseline uses the parameters in Section \ref{sec:imp_8}: a filter radius of $4h$, a SIMP exponent increase of $0.1$ every $20$ iterations. We vary one parameter at a time relative to this baseline. The resulting designs are shown in Figure \ref{fig:numstudy_iso} for the isotropic material and in Figure \ref{fig:numstudy_aniso} for the anisotropic material.

\paragraph{Filter radius.}
The filter radius in the topology optimization scheme is a numerical (or design) parameter, hence there is not an optimal filter. This filtering introduces an effective regularization length, and can be selected in relation to the smallest manufacturing feature size. Increasing it to $1.5\times 4h$ produces the same general topology, but with smoother and larger features (Case F1 in Figure \ref{fig:numstudy_iso},\ref{fig:numstudy_aniso}). 
Decreasing it to $1.5\times 4h$ produces very similar topology in the isotropic case but did not converge to a meaningful design in the anisotropic case (Case F2). In general, a smaller filter might result in non-convergence because it allows sharper element density variations and thinner load paths, which can produce mechanically unstable intermediate designs and the forward problem becomes poorly conditioned. 

\paragraph{SIMP exponent.}
Since optimal design is an ill-posed problem, relaxation of the problem needs to be performed by introducing SIMP interpolation and SIMP exponent ($p$). We aim for a solid-void design, so we follow common practice to increase the SIMP exponent $(p)$ from 1 to 3 to penalize gray values. We investigate the results for: SIMP rate = $3\times$, which corresponds to an increase of $0.1$ every $60$ iterations. We also consider SIMP rate = $0.75\times$, which corresponds to an increase of $0.1$ every $15$ iterations. For both materials, both rates result in topologies similar to baseline. (Case S1, S2).

\begin{figure}
  \centering
  \includegraphics[width=0.3\textwidth]{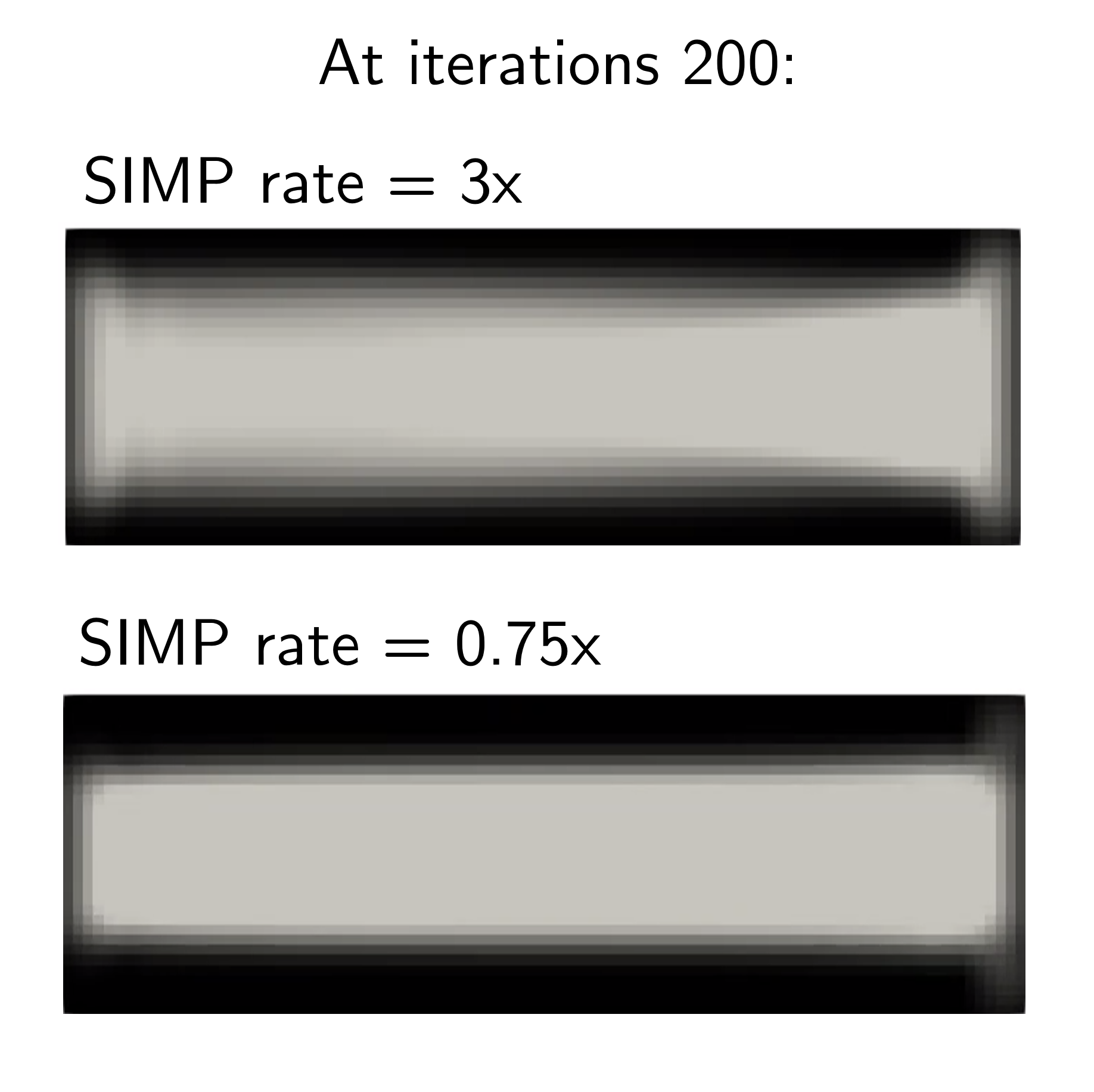}
  \caption{Isotropic hyperelastic material: design at iteration $200$ for a SIMP rate of $3\times$ (top) and $0.75\times$ (bottom) the baseline. The SIMP rate controls how quickly the design converges to a solid--void layout.}
  \label{fig:simprate}
\end{figure}

In general, the SIMP rate controls the rate at which the design converges to a solid-void design (Figure \ref{fig:simprate}). Increasing the SIMP exponent too rapidly can quickly penalize intermediate densities and those regions become much softer before the design reorganizes, which can cause failure in the forward problem. 

\paragraph{Noise levels.}
Finally, we vary the magnitude of the noise covariances. The baseline noise covariances are $\Gamma_f = (0.01t)^2\mathbf{I}$ and $\Gamma_u = (10^{-5})^2\mathbf{I}$ for isotropic material (Section \ref{sec:iso_8}), or $\Gamma_u = (10^{-6}\times t)^2\mathbf{I}$ for anisotropic material (Section \ref{sec:fiberimg_8}). We consider larger displacement and smaller force noise in Case N1, smaller displacement and larger force noise in Case N2, and the same displacement and smaller force noise in case N3. For both materials, the main geometric features of the baseline design are preserved in all three cases.

In summary, across all tested perturbations, these resulting designs have similar overall geometric features and indicate that the proposed procedure is robust to the considered variations in the prior, initial geometry, noise level, and numerical parameters.

\section{Conclusion}

We have developed an approach to design specimens to enable efficient and accurate recovery of material parameters from experiments.  We pose the problem of material parameter identification as a Bayesian inverse problem, and cast the problem of specimen design as a D-optimal experimental design problem solved using topology optimization. We demonstrate our approach on isotropic and anisotropic hyperelastic materials, and show that the optimized specimens increase the expected information gain and provide more accurate parameter inference compared to conventional or non-optimized designs. 

The results in this work suggest that direct image-based inference is valuable to obtain high-fidelity constitutive models, while displacement-based data may provide a more stable and practical route for OED, leading to more reliable designs. For model parameter inference, raw images contain rich, fine-scale information and, when used within the image-to-constitutive integrated framework (e.g.,\cite{wihardja2025constitutive}), avoid the intermediate step to obtain displacements. However, using raw images to guide optimal experimental design is challenging since design sensitivity depends on the image intensities, and differentiating them leads to noisy gradients that poorly influence the topology optimization.

We close by discussing future extensions building on this framework. A natural extension is to consider history-dependent materials, such as plasticity and rate-dependent soft materials. In these complex systems, the constitutive response depends not only on the current deformation but also on internal states and loading history, and is thus more challenging to characterize. This added complexity opens interesting directions.  First, the design problem should consider not only the specimen topology, but also the loading path and rate to sufficiently explore relevant evolution laws. Second, this increased complexity calls for the development of an adaptive and fast data assimilation framework that continuously integrates experiments with model inference. In particular, as we gain more knowledge about the material history from experiments (e.g., evolution of yield surface), we can perform follow-up optimal experiments using this updated knowledge in real time. This necessitates a scalable Bayesian OED framework. For strongly nonlinear, history-dependent materials, the adjoint-based gradient methods presented here may become prohibitively expensive, making gradient-free ensemble methods, such as ensemble Kalman inversion or ensemble Kalman filtering, attractive for coupling with the Bayesian optimal experimental design framework here.   

Beyond extension to more complex materials, an important consideration is robustness to real experimental conditions. In practice, in any fabrication and experiment, divergence from design to execution is inevitable. This challenge can be addressed in two interesting ways. First, robustness can be incorporated directly into the design problem by optimizing specimens that remain informative under expected manufacturing and experimental variability (e.g., printing defects, material heterogeneity). Second, robustness can be handled sequentially, where after each experiment, the actual fabricated geometry, noise, and measured response can be assimilated into the entire model update, and the next design can be chosen based on the accumulated data and observed deviations. Such an integrated closed-loop framework would make the proposed approach more practical for real experiments to capture materials with complex physics, while further reducing the number of tests required for high-fidelity constitutive model discovery in solid mechanics.

This work focuses on parameter identification with the form of the constitutive equation chosen {\it a priori}.  This presupposes the adequacy of the constitutive framework to describe the complexity of the phenomenon understudy.  The Bayesian framework can be extended to evaluate the adequacy of the constitutive equation through Bayesian model selection.  The idea is to introduce a prior probability over candidate models, and this remains a difficult challenge.  If one were able to do so, say by parametrizing the class of candidate models, then we can use the marginal likelihood as the model evidence.  This defines the posterior probability that enters the design objective, or the expected information gain.  Unfortunately, the current closed-form log-det EIG approximation does not apply in this setting.  Therefore additional development is required here.

The framework and ideas developed in this work are applicable to three space dimensions.  However, the present implementation is limited to the two-dimensional plane-stress setting motivated by the established experimental setup where thin specimens are used and loaded in a Universal Testing Machine.  Three dimensions is  relevant for X-ray tomography and confocal microscopy techniques that are being developed, but these require different formulation of the parameter identification obtjective (Section \ref{sec:param_8}).   Further, the fabrication of specimens with complex geometry is extremely difficult in three dimensions, and one would have to incorporate manufacturing constraints into the formulation.

\paragraph{Acknowledgement}
We are delighted to acknowledge useful discussions with Andrew M.\ Stuart.  We gratefully acknowledge the financial support of the Caltech MCE Center on AutoFab and the Office of Naval Research through MURI grant Number N00014-23-1-2654.

\paragraph{Author contributions: CRediT}
Adeline Wihardja: Conceptualization, Formal analysis, Investigation, Methodology, Software, Validation, Visualization, Writing -- original draft, Writing -- review and editing.
Kaushik Bhattacharya:  Conceptualization, Formal analysis, Funding acquisition, Investigation, Methodology, Supervision, Validation, Visualization, Writing -- original draft, Writing -- review and editing.


\section*{Appendix}
\appendix
\section{Derivation for sensitivity and adjoint PDEs} \label{app:adj_88}
We seek to find the gradient of $\mathcal{O}_\text{EIG}$ with respect to $\rho$.  The details of this gradient calculation to obtain the sensitivity (\ref{eq:oedsens_8}) and the adjoints (\ref{eq:betaP_8},\ref{eq:lambdaadj_8}) are given here. Given any $\lambda, \beta_P \in {\mathcal U}$, we use the forward problems (\ref{eq:res1_8} and (\ref{eq:res2_8}) to rewrite our objective as
\begin{align}
 \mathcal{O}^*(\rho,u, u_P)  &= {-\mathcal{O}_{EIG}}(\rho,u, u_P)  + \int_\Omega (\rho^p \ W_F(\nabla u, P_0))  \cdot \nabla \lambda\, d\Omega \\
    & \quad \quad + \int_\Omega  \left(  \rho^p \left(W_{FP} (\nabla u, P_0)  +  W_{FF} (\nabla u, P_0) \nabla u_P \right) \right)\cdot \nabla \beta_P \, d\Omega.
\end{align}
For the rest of the calculations in this section, we define 
\begin{equation}
    u_{P_m} :=\frac{\partial u}{\partial P_m}, \quad v_{P_m} :=\delta u_{P_m}, \quad v:=\delta u,
\end{equation}
and define
\begin{equation}
    H_{mn} = \int_0^T\Bigg(\int_\Omega \rho^{2p} u_{P_m}^T\,\,\Gamma_u^{-1}\,\,u_{P_n} \,d\Omega + J_m^T\,\,\Gamma_f^{-1}\,\,J_n \Bigg) dt,
\end{equation}
where $m,n$ denote parameter indices. In general, $H$ is a finite-dimensional matrix depending on the number of material parameters $P\in\mathbb{R}^P$ that are considered, i.e., $H\in\mathbb{R}^{P\times P}$. For the rest of this section, we explicitly denote $H$ as $H_{mn}$. We follow Einstein summation over the parameter indices. The EIG is 
\begin{equation}
    \mathcal{O}_{EIG} = \frac{1}{2}\log\det(I+C_0H), \text{ $H$ evaluated with $P_0$ for a given design $\rho$}.
\end{equation}
Thus,
\begin{equation}
    \delta\mathcal{O}_{EIG} = \frac{1}{2}\left[C_0(I+C_0H)^{-T}\right]_{mn}\delta H_{mn}.
\end{equation}
and
\begin{equation}
    A_{mn} :=  \left[ \frac{1}{2} C_0(I+C_0H)^{-T}\right]_{mn} = \left[\frac{1}{2}C_\text{post}\right]_{mn}
\end{equation}
Now, perturb $\rho$ in the direction $\mu$, such that $\rho \mapsto \rho+ \varepsilon \mu$,  $u \mapsto u + \varepsilon v$, $u_{P_m}\mapsto  u_{P_m} + \varepsilon v_{P_m}$.  
The full variation of $\mathcal{O}^*$ with respect to $\rho$ in the direction $\mu$ is given by
\begin{equation}
\langle \delta_\rho \mathcal{O}^*, \mu \rangle := \left.\frac{d}{d\varepsilon} {\mathcal O}^*(\rho + \varepsilon \mu, u + \varepsilon v, u_P + \varepsilon v_P) \right|_{\varepsilon = 0}.
\end{equation}

\noindent Carrying out the calculation:
\begin{equation} 
\begin{aligned}
&\langle \delta_\rho \mathcal{O}^*, \mu \rangle = \\
& -A_{mn}  \Biggr(\int_0^t \int_\Omega 2p \rho^{p-1} \,u_{P,m}^T \,\, \Gamma_u^{-1} \,\,\rho^p \, u_{P,n} \, d\Omega dt\Biggr)\mu \\   
& -A_{mn} \int_0^t \int_{\Omega}  \rho^{2p} \, u_{P_m}^T \,\, \Gamma_u^{-1} \,\, v_{P_n} + \rho^{2p} \, v_{P_m}^T \,\, \Gamma_u^{-1} \,\, u_{P_n} \, d \Omega dt  \\  
&\\
&- A_{mn}  \int_0^t \bigg(\int_{\partial_u \Omega} \left( \left( W_{FF} \nabla v_{P_m} \right)\hat{n}\right) \cdot e_f\,  d A \, \biggr)^T \,\, \Gamma_f^{-1} \,\, J_{f_n} +  J_{f_m}^T \,\, \Gamma_f^{-1} \bigg(\int_{\partial_u \Omega} \left( \left( W_{FF} \nabla v_{P_n} \right)\hat{n}\right) \cdot e_f\,  d A \, \biggr)\, dt   \\
&-A_{mn} \int_0^t   J_{f_m}^T \,\, \Gamma_f^{-1}  \Biggr( \int_{\partial_u \Omega} \left( \left( W_{FFF} \nabla u_{P_n} + W_{FFP_n} \right) \nabla v \right)\hat{n}\cdot e_f \,\, dA \Biggr) \, dt\\
&-A_{mn} \int_0^t   \Biggr( \int_{\partial_u \Omega} \left( \left( W_{FFF} \nabla u_{P_m} + W_{FFP_m} \right) \nabla v \right)\hat{n}\cdot e_f\,\, dA \Biggr)^T  \,\, \Gamma_f^{-1} \,\, J_{f_n}  \, dt\\
&\\
&+ \int_\Omega   \rho^p \, W_{FF} \nabla v  \cdot  \nabla \lambda \, d\Omega  \\
&+ \int_\Omega \Big( p\rho^{p-1} \, W_F  \cdot \nabla \lambda \Big) \mu \, d\Omega  \\
&\\
& + \int_\Omega   p\rho^{p-1}  \Big(\, W_{FP_m}  +  W_{FF}  \nabla u_{P_m} \Big) \cdot \nabla \beta_{P_m}  \mu \, d\Omega\\
&+ \int_\Omega \rho^p \bigg(W_{FFF} \nabla v \nabla u_{P_m}  + W_{FFP_m} \nabla v + W_{FF} \nabla v_{P_m}\bigg)\cdot \nabla \beta_{P_m}  \, d\Omega
\end{aligned}
\end{equation}
Grouping terms with $\mu$, we obtain the sensitivity to be
\begin{equation} 
\begin{aligned}
&-A_{mn}  \Biggr(\int_0^t \int_\Omega 2p \rho^{p-1} \,u_{P,m}^T \,\, \Gamma_u^{-1} \,\,\rho^p \, u_{P,n} \, d\Omega dt\Biggr)\mu \\   
&+ \int_\Omega \Big( p\rho^{p-1} \, W_F  \cdot \nabla \lambda \Big) \mu \, d\Omega  \\
& + \int_\Omega   p\rho^{p-1}  \Big(\, W_{FP_m} +  W_{FF} \nabla u_{P_m} \Big) \cdot \nabla \beta_{P_m}  \mu \, d\Omega.
\end{aligned}
\end{equation}
Grouping terms of $v_{P_m}$ and setting to zero,
\begin{equation}
\begin{aligned}
0 = &-A_{mn} \int_0^t \int_{\Omega}  \rho^{2p} \, v_{P_m}^T \,\, \Gamma_u^{-1} \,\, u_{P_n} \, d \Omega dt  \\
&-A_{nm} \int_0^t \int_{\Omega}  \rho^{2p} \, u_{P_n}^T \,\, \Gamma_u^{-1} \,\, v_{P_m} \, d \Omega dt  \\
&- A_{mn}  \int_0^t \biggr(\int_{\partial_u \Omega} \left( \left( W_{FF} \nabla v_{P_m} \right)\hat{n}\right) \cdot e_f\,  d A \, \biggr)^T \,\, \Gamma_f^{-1} \,\, J_{f_n} \, dt   \\
&- A_{nm}  \int_0^t  J_{f_n}^T \,\, \Gamma_f^{-1} \bigg(\int_{\partial_u \Omega} \left( \left( W_{FF} \nabla v_{P_m} \right)\hat{n}\right) \cdot e_f\,  d A \, \biggr)\, dt   \\
&+ \int_\Omega \rho^p \bigg( W_{FF} \nabla v_{P_m}\bigg)\cdot \nabla \beta_{P_m}  \, d\Omega \quad \quad \forall v_{P_m} \in \mathcal{U},
\end{aligned} 
\end{equation}
furnishes the adjoint equation A1 for $\beta_{P_m}$ for each parameter $m$.\\

\noindent Grouping terms of $v$ and setting to zero,
\begin{equation}
\begin{aligned}
0 = &- A_{mn} \int_0^t   J_{f_m}^T \,\, \Gamma_f^{-1}  \Biggr( \int_{\partial_u \Omega} \left( \left( W_{FFF} \nabla u_{P_n} + W_{FFP_n} \right) \nabla v \right)\hat{n}\cdot e_f \,\, dA \Biggr) \, dt\\
&-A_{mn} \int_0^t   \Biggr( \int_{\partial_u \Omega} \left( \left( W_{FFF} \nabla u_{P_m} + W_{FFP_m} \right) \nabla v \right)\hat{n}\cdot e_f\,\, dA \Biggr)^T  \,\, \Gamma_f^{-1} \,\, J_{f_n}  \, dt\\
&+ \int_\Omega   \rho^p \, W_{FF} \nabla v  \cdot  \nabla \lambda \, d\Omega  \\
&+ \int_\Omega \rho^p \bigg(W_{FFF} \nabla v \nabla u_{P_m}  + W_{FFP_m} \nabla v \bigg)\cdot \nabla \beta_{P_m}  \, d\Omega \quad \quad \forall v\in\mathcal{U},
\end{aligned}
\end{equation}
results in the adjoint equation A2 for $\lambda$.
We note that all the derivatives above, i.e., $W_{F},W_{FF},W_{FFF},W_{FP},W_{FFP}$ are all evaluated at $(\nabla u, P_0)$, and density $\rho$ is the proposed design.

\end{document}